\documentclass[iop]{emulateapj}
\usepackage[usenames,dvipsnames]{xcolor}
\usepackage{multirow}

\newcommand\cii{[C~{\sc ii}]~}

\shorttitle{A non-corotating gas component}
\shortauthors{Tadaki et al.}

\begin{document}


\title{A non-corotating gas component in an extreme starburst at $z=4.3$}


\author{Ken-ichi Tadaki\altaffilmark{1},
Daisuke Iono\altaffilmark{1,2},
Min S. Yun\altaffilmark{3},
Itziar Aretxaga\altaffilmark{4},
Bunyo Hatsukade\altaffilmark{5},
Minju M. Lee\altaffilmark{6},
Tomonari Michiyama\altaffilmark{1},
Kouichiro Nakanishi\altaffilmark{1,2},
Toshiki Saito\altaffilmark{7},
Junko Ueda\altaffilmark{1}, and
Hideki Umehata\altaffilmark{5,8}
}


\affil{\altaffilmark{1} National Astronomical Observatory of Japan, 2-21-1 Osawa, Mitaka, Tokyo 181-8588, Japan; tadaki.ken@nao.ac.jp}
\affil{\altaffilmark{2} Department of Astronomical Science, SOKENDAI (The Graduate University for Advanced Studies), Mitaka, Tokyo 181-8588, Japan}
\affil{\altaffilmark{3} Department of Astronomy, University of Massachusetts, Amherst, MA 01003, USA}
\affil{\altaffilmark{4} Instituto Nacional de Astrofisica, Opticay Electronica (INAOE), Puebla, Mexico}
\affil{\altaffilmark{5} Institute of Astronomy, Graduate School of Science, The University of Tokyo, 2-21-1 Osawa, Mitaka, Tokyo 181-0015, Japan}
\affil{\altaffilmark{6} Max-Planck-Institut f\"ur extraterrestrische Physik (MPE), Giessenbachstr., D-85748 Garching, Germany}
\affil{\altaffilmark{7} Max-Planck Institute for Astronomy, K\"onigstuhl, 17 D-69117 Heidelberg, Germany}
\affil{\altaffilmark{8} RIKEN Cluster for Pioneering Research, 2-1 Hirosawa, Wako-shi, Saitama 351-0198, Japan}



\begin{abstract}
We report the detection of a non-corotating gas component in a bright unlensed submillimeter galaxy at $z=4.3$, COSMOS-AzTEC-1, hosting a compact starburst.
ALMA 0.17 and 0.09 arcsec resolution observations of \cii emission clearly demonstrate that the gas kinematics is characterized by an ordered rotation.
After subtracting the best-fit model of a rotating disk, we kinematically identify two residual components in the channel maps.
Both observing simulations and analysis of dirty images confirm that these two subcomponents are not artificially created by noise fluctuations and beam deconvolution.
One of the two has a velocity offset of 200 km s$^{-1}$ and a physical separation of 2 kpc from the primary disk and is located along the kinematic minor axis of disk rotation.
We conclude that this gas component is falling into the galaxy from a direction perpendicular to the disk rotation.
The accretion of such small non-corotating gas components could stimulate violent disk instability, driving radial gas inflows into the center of galaxies and leading to formation of in-situ clumps such as identified in dust continuum and CO.
We require more theoretical studies on high gas fraction mergers with mass ratio of 1:$>10$ to verify this process.
\end{abstract}





\section{Introduction}

Bright submillimeter galaxies (SMGs) are intensively forming stars with a rate of $1000~M_\odot$yr$^{-1}$, except for strongly-lensed objects. 
The dust continuum emission is compact with a half-light radius of $\sim$1 kpc \citep[e.g.,][]{2015ApJ...799...81S, 2015ApJ...810..133I, 2016ApJ...833..103H}, which corresponds to the size of a bulge in massive quiescent galaxies at $z\sim2$ and giant elliptical galaxies at $z=0$ \citep[e.g.,][]{2005ApJ...626..680D,2006MNRAS.373L..36T,2011ApJ...739L..44D,2015ApJ...813...23V}.
These findings suggest an evolutionary link between bright SMGs and compact quiescent galaxies at $z\sim2$ \citep{2014ApJ...782...68T} although it is not necessarily the case in faint SMGs with a flux density of $<$3.5 mJy at 850 $\mu$m \citep{2019arXiv190910540V}.
The star formation rate surface density in the central 1--2 kpc region exceeds 100 $M_\odot$yr$^{-1}$kpc$^{-2}$ \citep[e.g.,][]{2008ApJ...688...59Y, 2014PhR...541...45C, 2018Natur.560..613T}.
Understanding the physical mechanism triggering such an extreme starburst in early Universe is a main topic in this paper.

In the hierarchical structure formation scenario, galaxies build up their stellar mass, morphology, and angular momentum through multiple-mergers \citep[e.g.,][]{2013MNRAS.428.2529H}.
Major mergers with a mass ratio from 1:1 to 1:4 drive gas inflows into the center, leading to a nuclear starburst as well as a feeding a supermassive black hole in the center of a galaxy \citep[e.g.,][]{2006ApJS..163....1H}.
In local Universe, extreme starburst galaxies with a total infrared luminosity of $L_\mathrm{IR}>10^{11.8}~L_\odot$ are all associated with an equal-mass companion with a separation of less than 10 kpc \citep[e.g.,][]{1988ApJ...325...74S, 2016ApJ...825..128L}.
Although some CO or \cii line observations of high-redshift SMGs identify gas-rich companions with a separation of $30-200$ kpc \citep[e.g.,][]{2008ApJ...680..246T, 2009ApJ...694.1517D, 2012MNRAS.425.1320I, 2013ApJ...772..137I, 2014ApJ...796...84R}, 
there is not clear evidence that they are in a late-stage of major mergers or in the final coalescence where the star forming activity is the most enhanced. 
Parsec-resolution hydrodynamical numerical simulations demonstrate that major mergers of galaxies with a high gas fraction of 60\% are less efficient at producing starbursts compared to mergers with a low gas fraction of 10\%, expected at low-redshift \citep{2017MNRAS.465.1934F}.
Even if high-redshift SMGs experience a major merger, it could not necessarily trigger an extreme starburst.

Minor mergers with a mass ratio of 1:10 are expected to happen more frequently \citep[e.g.,][]{2004ApJ...617L...9L,2009ApJ...697.1971J,2009MNRAS.394.1713K} and contribute to enhancements in star formation activity \citep[e.g.,][]{2014MNRAS.437L..41K,2016A&A...587A..24S}.
The accretion of small satellites onto massive galaxies could also affect the dynamical condition and cause compaction of the gas disk if the gas fractions of the companions are high \citep[e.g.,][]{2009Natur.457..451D, 2015MNRAS.450.2327Z}.
However, late-stage minor mergers are poorly explored in observations of high-redshift SMGs because it is difficult to identify small companions at a distance of less than 10 kpc ($<$1.\arcsec5) due to limitations of sensitivity and spatial resolution in submillimeter/millimeter observations.

\begin{figure*}[th!]
\begin{center}
\includegraphics[scale=1.0]{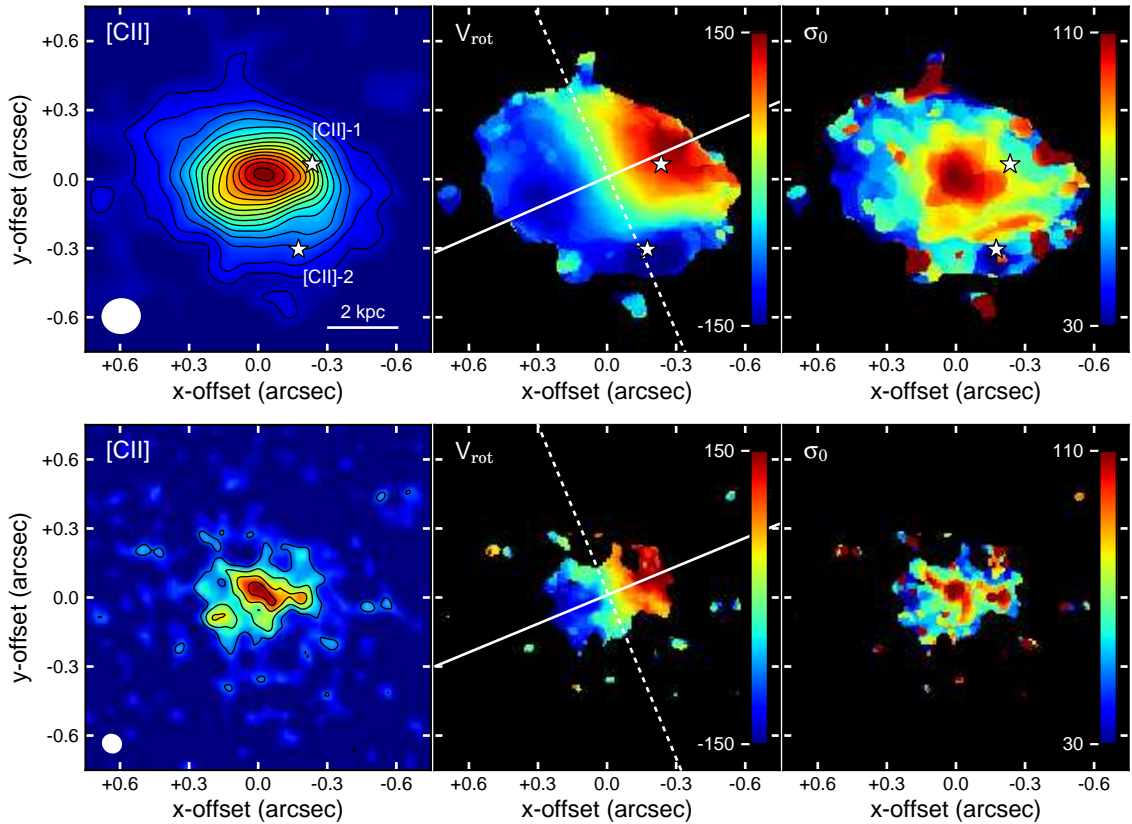}
\end{center}
\caption{
The low-resolution (0.\arcsec17; top panels) and high-resolution (0.\arcsec09; bottom panels) ALMA images of \cii line emission.
From left to right, we show the velocity-integrated flux maps, the line-of-sight velocity maps and local velocity dispersion maps.
White solid and dashed lines show the kinematic major and minor axis of disk rotation, respectively.
White stars mark the positions of \cii subcomponents (Section \ref{sec;cii_clump}).
Contours are plotted every 2$\sigma$ from 3$\sigma$.
}
\label{fig;map}
\end{figure*}

Recent high-resolution observations have revealed off-center gas clumps in SMGs and dusty star-forming galaxies \citep[e.g.,][]{2012ApJ...760...11H, 2014MNRAS.442..558A, 2016ApJ...829L..10I, 2019ApJ...876..130H, 2019ApJ...882..107R}.
The gas mass surface densities are extremely high $\Sigma_\mathrm{gas}\sim10^4~M_\odot$ pc$^{-2}$ in the central 1--2 kpc region and then the self-gravity of the gas overcomes the internal pressure due to stellar-radiation feedback with Toomre $Q$ parameters of $Q<1$ \citep[e.g.,][]{2011ApJ...733..101G, 2014A&A...565A..59D, 2015ApJ...806L..17S, 2018Natur.560..613T, 2019ApJ...870...80L}.
Off-center clumps are therefore expected to be formed through the gravitational instability of the dense gas disk.
On the other hand, they potentially have an ex-situ origin such as late-stage minor mergers.
In this paper, we report the detection of a non-corotating subcomponent in a bright SMG at $z=4.3$, COSMOS-AzTEC-1, from deep and high-resolution Atacama Large Millimeter/submillimeter Array (ALMA) observations of 860 $\mu$m continuum and \cii line emission.

\section{Data}

We use data-sets from ALMA Band-7 observations of COSMOS-AzTEC-1 with a central frequency of 350 GHz (860 $\mu$m), conducted in two array configurations (\citealt{2016ApJ...829L..10I}, \citealt{2019ApJ...876....1T}).
The compact and extended configuration observations cover the baseline lengths of 15 m -- 2.5 km and 178 m -- 14.6 km, respectively.
The integration time is $\sim$30 minutes in both configurations.
We calibrate the data in the standard manner using {\tt CASA} \citep{2007ASPC..376..127M}.
There is no systematic offset in the amplitude of the vector-averaged visibilities between the compact and extended configuration data (Appendix A).

First, we make a high-resolution 860 $\mu$m continuum map by combining the two configuration data.
We use only two spectral windows because the other two detect \cii 158 $\mu$m and OH 163 $\mu$m emission lines.
We decrease the absolute visibility weights of the compact data by 0.2, and then combine the two configuration data  in the visibility plane using the {\tt CASA/concat} task.
Cleaning with weighting of {\tt robust}=+2.0 and {\tt uvtaper}=0.\arcsec05 results in a spatial resolution of 0.\arcsec09$\times$0.\arcsec09 with reasonable sidelobe levels.
We clean the map down to the 1$\sigma$ level in a circular mask with a diameter of 1\arcsec using the {\tt CASA/tclean} task.
We also adopt {\tt Multi-Scale Clean} algorithm \citep{2008ISTSP...2..793C} to detect extended uncleaned emission in the smoothed residual images. 
We estimate the noise level by calculating the root mean square and the standard deviation about the mean in the annular region between radii of 1 arcsec and 9 arcsec, corresponding to the primary beam size of the clean map. 
These two measurements are identical, resulting in the noise level of 46 $\mu$Jy beam$^{-1}$.

\begin{figure*}
\begin{center}
\includegraphics[scale=1.0]{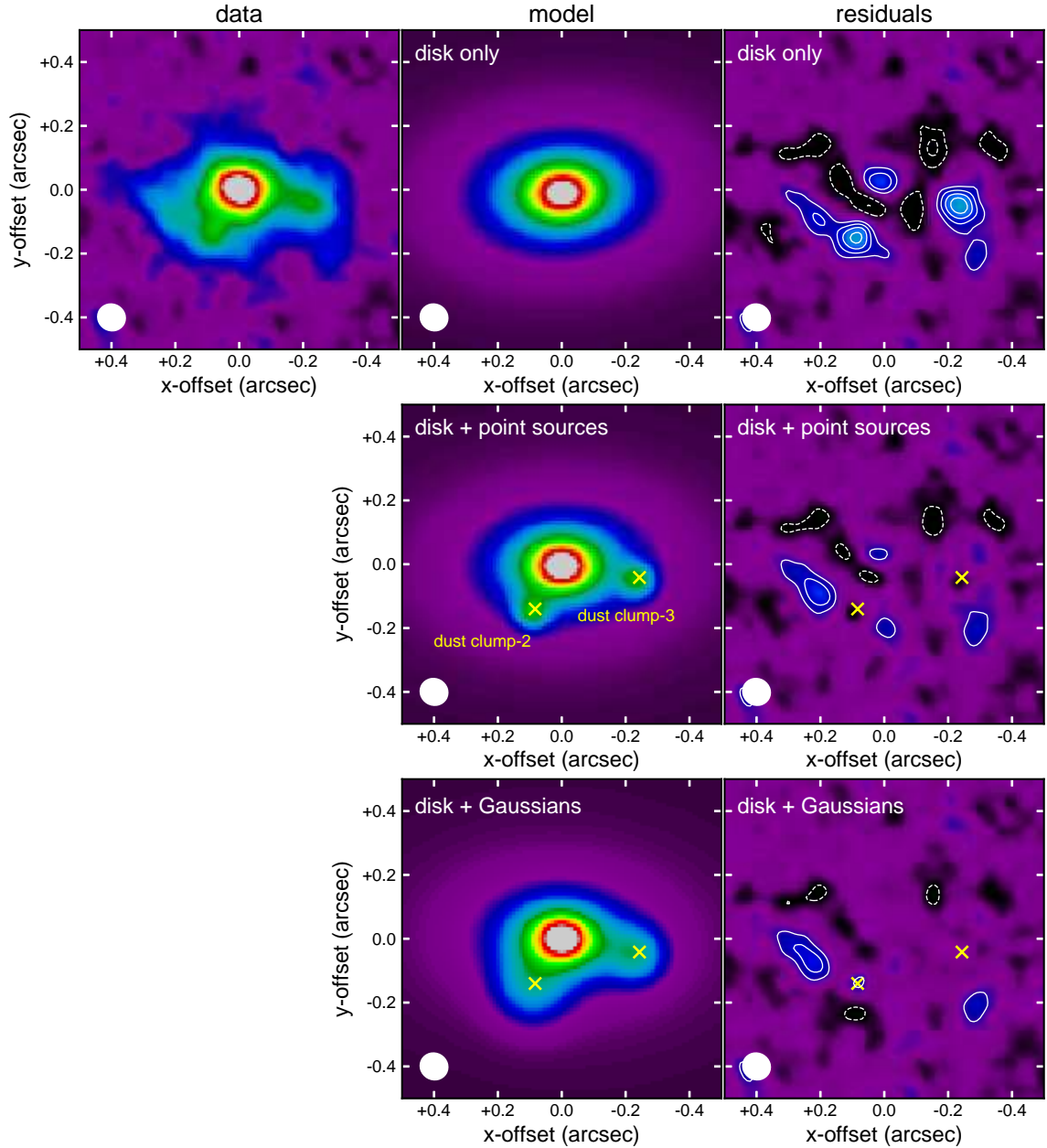}
\end{center}
\caption{
The high-resolution (0.\arcsec09) ALMA images of 860 $\mu$m continuum emission in COSMOS-AzTEC-1.
From left to right, we show the clean map, the best-fit model and the residuals after subtracting the primary disk component. 
The image size is 1\arcsec$\times$1\arcsec.
We use three models: single S$\acute{\mathrm{e}}$rsic component (top), S$\acute{\mathrm{e}}$rsic + two point sources (middle), and S$\acute{\mathrm{e}}$rsic + two Gaussians (bottom).
White contours show the -5,-3, 3, 5, 7, 9$\sigma$ levels in the residual maps.
Yellow crosses mark the positions of two dust clumps.
}
\label{fig;dustmap}
\end{figure*}

Next, we make low- and high-resolution clean cubes with a channel width of 30 km s$^{-1}$ for studying the \cii emission line in COSMOS-AzTEC-1.
We simply combine the two configuration data without decreasing the absolute visibility weights and create a low-resolution cube with the parameters of {\tt robust}=+2.0 and {\tt uvtaper}=0.\arcsec1, resulting in a spatial resolution of 0.\arcsec$17\times$0.\arcsec16.
The noise levels in the low- and high-resolution cubes are 0.40 mJy beam$^{-1}$ per channel and 0.45 mJy beam$^{-1}$ per channel, respectively.
We make \cii maps of velocity-integrated flux, velocity field, and velocity dispersion in the velocity range between -240 and +240 km s$^{-1}$ using the {\tt CASA/immoments} task (Figure \ref{fig;map}).
Both of the low- and high-resolution velocity maps show a clear sign of rotation with a monotonic gradient from southeast to northwest.

\section{Analysis}

Previous 0.\arcsec05-resolution 860 $\mu$m continuum map reveals that COSMOS-AzTEC-1 is composed of a central starburst disk and multiple off-center clumps \citep{2016ApJ...829L..10I}.
We identify clump candidates by using the new dust continuum map and \cii line cube.
In both cases, we first fit disk models to the data and then search for the residual emission in the model-subtracted images.

\subsection{Clump identification in dust continuum emission}
\label{sec;dust_clump}

To extract the primary disk component, 
we fit single S$\acute{\mathrm{e}}$rsic models to the high-resolution 860 $\mu$m map using the {\tt GALFIT} code \citep{2010AJ....139.2097P}.
We use a clean Gaussian beam as a point-spread function for deconvolution.
There are seven free parameters of the models: centroid position ($x$, $y$), 860 $\mu$m continuum flux density $S_{860}$, half-light radius $R_{1/2}$, S$\acute{\mathrm{e}}$rsic index $n$ ($n=0.5$ for Gaussian, $n=1$ for an exponential profile, $n=4$ for de Vaucouleurs profile), minor-to-major axis ratio $q$, and position angle while the sky value is fixed to be zero.
We obtain the best-fit model of the primary disk component with fitting errors of 1\% or less (Figure \ref{fig;dustmap}).
On the other hand, simulations of ALMA observations indicate that there are much larger systematic and random errors in the measurements (3--10 \%; see Appendix B).
We take into account these errors but note that they do not include uncertainties due to differences between real galaxies and idealized profiles \citep{2010AJ....139.2097P}.
We derive $S_{860}=17.50\pm0.74$ mJy, $R_{1/2}=1.22\pm0.07$ kpc, $n=1.26\pm0.09$ and $q=0.64\pm0.02$, which are consistent with the previous results that the morphology of dust emission is characterized by a compact exponential disk \citep{2016ApJ...833..103H, 2018ApJ...861....7F}.
In the residual map after subtracting the S$\acute{\mathrm{e}}$rsic disk, we detect the emission at $5\sigma$ in several regions as well as negative $5\sigma$ emission (Figure \ref{fig;dustmap}).
\cite{2016ApJ...829L..10I} previously identified 11 clumps at above 4$\sigma$ and 3 out of 11 (dust clump-1, -2, -3) are detected at above 5$\sigma$.
The two high S/N clumps (dust clump-2 and dust clump-3) are identified at $9.7\sigma$ and $10.6\sigma$ in the residual map from this analysis, respectively.
These bright clumps are likely to pull the central position of the best-fit models toward south, possibly causing the central residual emission and negative residuals around the center.
As the brightest dust clump (dust clump-1) is very close to the nuclei, it seems to be hard to isolate this clump even at 0.\arcsec09 resolution.
On the other hand, other 8 low S/N (4--5$\sigma$) clumps were not detected in the residual map.
The extended configuration data presented in \cite{2016ApJ...829L..10I} was only observed for a short $\sim$30 minutes on-source time and has relatively sparse $uv$ coverage especially at the longer baselines. 
This can possibly lead to low S/N artifacts in the map, and the true spatial distribution can only be revealed by obtaining a better $uv$ coverage with longer on-source integration time.
We caution that low S/N off-center clumps could be misidentified due to noise fluctuations on a smooth disk.
\cite{2016ApJ...833..103H} show by observing simulations that smooth exponential disks break up into a few clumps.
It is therefore important to subtract the smooth component before identifying clumps so that we can avoid artifacts associated to the noises.

To take into account of two bright off-center clumps, we remake the fitting using three component models, which consist of single S$\acute{\mathrm{e}}$rsic profile for the primary disk and two point sources for dust clump-2 and dust clump-3.
We derive $S_{860}=15.99\pm0.45$ mJy, $R_{1/2}=1.13\pm0.05$ kpc, $n=1.47\pm0.10$ and $q=0.66\pm0.02$ for the primary disk, $S_{860}=0.67\pm0.08$ mJy for dust clump-2 and $S_{860}=0.70\pm0.08$ mJy for dust clump-3.
The total flux density is $S_{860}=17.36\pm0.46$ mJy.
After subtracting the best-fit model, the residual map still shows a $6.7\sigma$ peak (Figure \ref{fig;dustmap}).
Simulations of ALMA observations demonstrate that $>6\sigma$ peaks are not artificially created by instrumental noise, contributions due to imperfect $uv$ sampling, and noise fluctuations on the underlying smooth disk (see Appendix B).
The residual component is likely to be a real substructure while it does not necessarily mean an independent clump from the disk.
Therefore we adopt a more strict criterion that residual emission is detected above 7$\sigma$ for identifying off-center clumps.

The deep high-resolution 860 $\mu$m map allows us not only to identify off-center clumps but also to constrain the spatial extent of the dust emission.
We here assume that off-center clumps are spatially extended and have a circular Gaussian shape, characterized by flux density and half-light radius. 
Then, we fit three component models (S$\acute{\mathrm{e}}$rsic plus two Gaussians) to the high-resolution 860 $\mu$m map (Figure \ref{fig;dustmap}).
We derive $S_{860}=13.62\pm0.69$ mJy, $R_{1/2}=1.04\pm0.08$ kpc, $n=1.58\pm0.14$ and $q=0.65\pm0.03$ for the disk, $S_{860}=2.09\pm0.31$ mJy and $R_{1/2}=0.54\pm0.06$ kpc for dust clump-2 and $S_{860}=1.38\pm0.18$ mJy and $R_{1/2}=0.38\pm0.05$ kpc for dust clump-3.
The total flux density is $S_{860}=17.09\pm0.78$ mJy.
The size of these dust clumps is much larger than that of giant molecular clouds in the Milky Way and nearby galaxies \citep[e.g.,][]{2008ApJ...686..948B,2015ARA&A..53..583H} but is comparable to that of clumps identified in local and high-redshift star-forming galaxies \citep[e.g.,][]{2005ApJ...627..632E,2012ApJ...752..111N,2017ApJ...839L...5F}.
Dust clump-2 and dust clump-3 contribute 12 percent and 8 percent to the total 860 $\mu$m flux density, respectively.

We find that the derived total 860 $\mu$m flux densities are consistent within the errors among the three models.
It would also make sense that clump flux densities in the Gaussian model become larger than those in the point source model because extended clumps get more flux density from the disk component. 
The extended model results in better fitting and well reproduces the overall distribution of dust continuum emission. 
On the other hand, the local distribution around the clumps more resembles the point source model, suggesting the possibility that the clumps are more compact than estimated from the Gaussian model. 
The eastern residual component makes fitting of the three component model difficult.
Adding more components to models is one way to more accurately assess the complex distribution whatever it is a part of the disk or an independent clump.

\begin{figure*}[t]
\begin{center}
\includegraphics[scale=1]{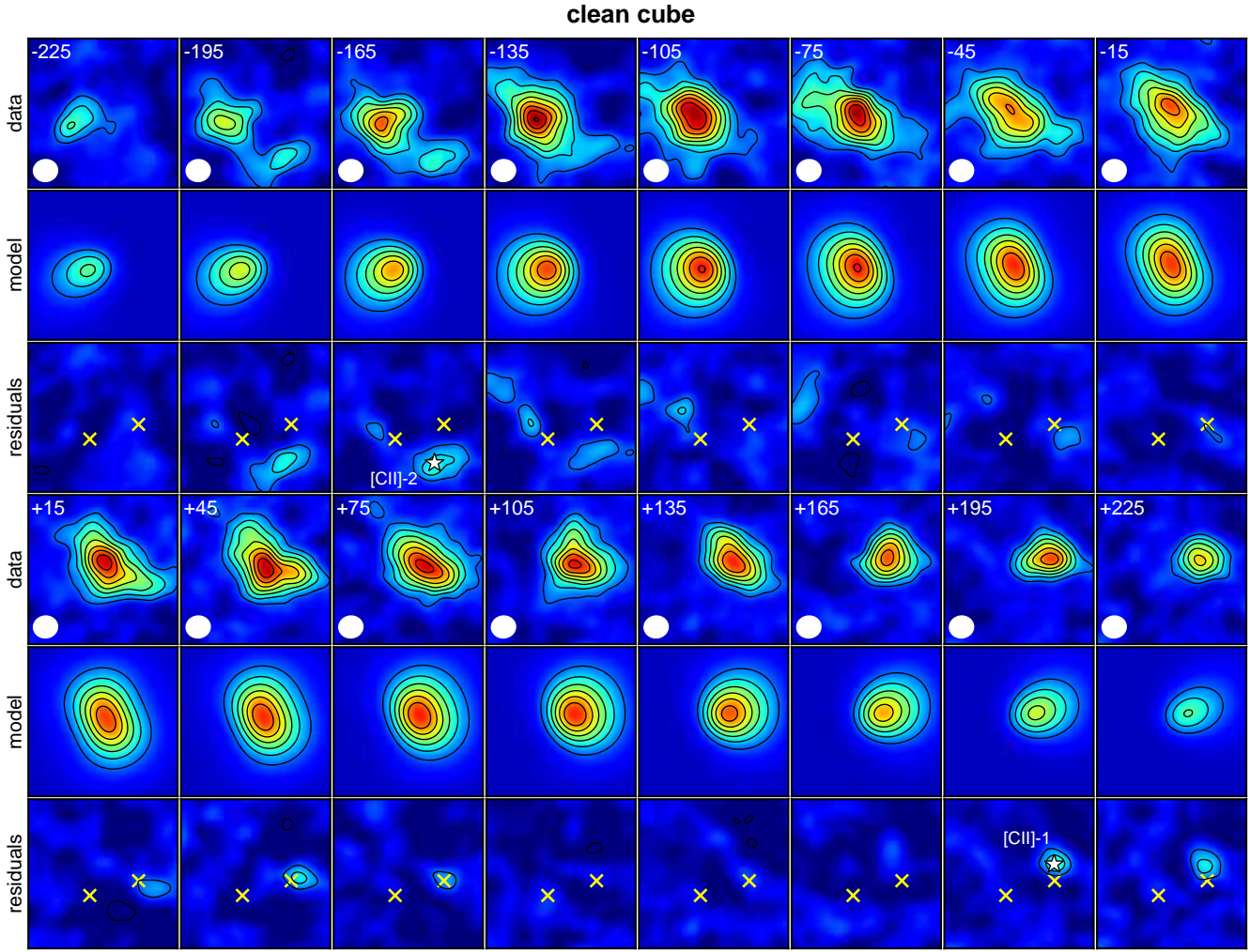}
\end{center}
\caption{
From top to bottom, the low-resolution (0.\arcsec17) \cii channel maps, the best-fit models and the residuals are shown in the velocity range from $-225$ km s$^{-1}$ to $+225$ km s$^{-1}$.
The image size is 1\arcsec$\times$1\arcsec.
Contours are plotted every 2$\sigma$ from 3$\sigma$.
Stars and crosses mark the positions of \cii subcomponents and dust clumps, respectively.
}
\label{fig;chmap}
\end{figure*}

\subsection{Residual components in \cii line emission}
\label{sec;cii_clump}

Next, we search for residual components in \cii emission after subtracting the disk component.
Figure \ref{fig;chmap} shows the low-resolution channel maps of the \cii emission.
The maximum signal-to-noise ratio is S/N=17 in a channel of -135 km s$^{-1}$ while the significance of the peak decreases to S/N=7 in the high-resolution channel maps.
Therefore, we use the low-resolution \cii cube for identifying residual components.
We fit kinematic models of a rotating disk to the \cii cube by using the {\tt Galpak3D} code \citep{2015AJ....150...92B}.
We assume a thick disk and an arctan rotation curve in the models. 
The models have 9 free parameters: centroid position, systemic velocity, line flux $Sdv$, maximum rotation velocity $V_\mathrm{max}$, local velocity dispersion $\sigma_0$ (not central velocity dispersion, but isotropic and constant one over the disk), half-light radius $R_{1/2}$, turnover radius, inclination and position angle. 
Disk models with these parameters are convolved with the clean beam and are fitted to the data cube using a Markov chain Monte Carlo (MCMC) algorithm.
We derive $Sdv=13.31\pm0.32$ Jy km s$^{-1}$, $V_\mathrm{max}=219\pm9$ km s$^{-1}$, $\sigma_0=74\pm2$ km s$^{-1}$ and $R_{1/2}=1.76\pm0.13$ kpc, where the systematic and random errors are taken into account on the basis of the simulations in a similar way as dust continuum observations (see Appendix B).
The kinematic parameters agree well with those derived from the previous fitting of a 0.3\arcsec-resolution \cii cube with a channel width of 50 km s$^{-1}$ \citep{2019ApJ...876....1T}, 
implying that measurements of maximum rotation velocity and local velocity dispersion do not require very high-resolution observations even in compact SMGs.
This is thanks to the fact that the \cii emission is relatively extended compared to the dust continuum and CO line emission \citep{2018ApJ...859...12G, 2019ApJ...876....1T}.
Given that \cii emission is completely resolved out in 0.\arcsec05-resolution cubes created from only the extended configuration data \citep{2016ApJ...829L..10I}, super-high-resolution observations of \cii emission are risky.
In the residual channel maps after subtraction of a smooth disk (Figure \ref{fig;chmap}), two subcomponents ([C~{\sc ii}]-1 and [C~{\sc ii}]-2) are detected at above $5\sigma$ in two adjacent channels, corresponding to the detection significance of $\sim7\sigma~(=5\sigma \times \sqrt{2})$ as the two channels are independent.
Both observing simulations and analysis of a dirty cube demonstrate that these residual components are not artificially created by noise fluctuations and beam deconvolution (see Appendix B and Appendix C).

\subsection{Properties of the off-center components}

\begin{figure}[!t]
\begin{center}
\includegraphics[scale=1.0]{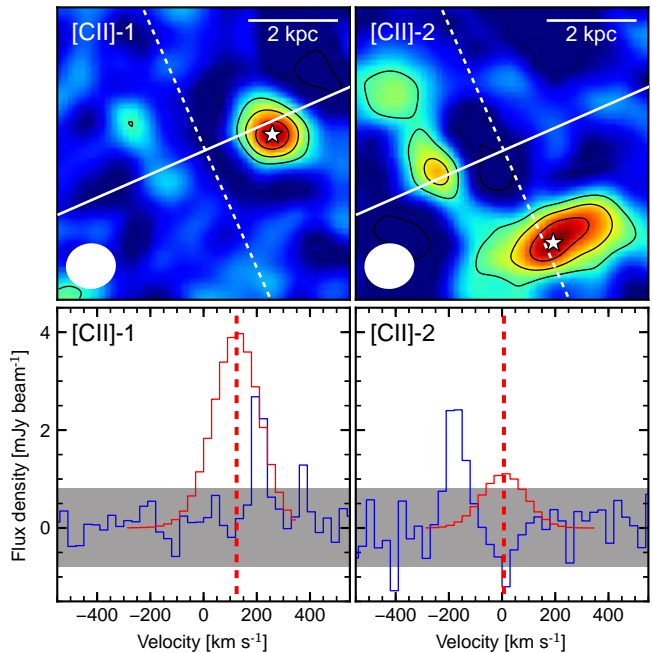}
\end{center}
\caption{
Spectra at the position of \cii subcomponents, extracted from the low-resolution cube.
Red and blue lines indicate the best-fit model as the primary disk component and the residuals between the data and the model, respectively.
Gray shaded regions denote the 2$\sigma$ level.
Red vertical lines show the coherent velocity of the disk component at the position of subcomponents.
The velocity-integrated residual maps are shown in the top panels.
The image size is 1\arcsec$\times$1\arcsec.
Contours are plotted every 2$\sigma$ from 3$\sigma$.
White solid and dashed lines show the kinematic major and minor axis of disk rotation, respectively (Figure \ref{fig;map}).
}
\label{fig;spe}
\end{figure}

Figure \ref{fig;spe} shows the \cii spectra of the best-fit disk models and the residuals at the position of the two subcomponents.
For [C~{\sc ii}]-1, the peak velocity of the residual emission is consistent within a range of $\sim100$ km s$^{-1}$ from the coherent velocity of the primary disk component, which is derived from single Gaussian fitting of the model spectrum.
Such a corotation is naturally expected if this component is formed through gravitational instability in the rotating starburst disk.
We note that a deviation from the idealized profile of the disk can possibly make such a corotating subcomponent, especially along the major axis of disk.
For example, an asymmetric disk or shocks could artificially produce a subcomponent in the residuals, which in that case would simply represent part of the disk emission. 
It requires deeper observations to distinguish whether or not \cii-1 truly represents a separate component from the overall disk rotation.
On the other hand, [C~{\sc ii}]-2 has a velocity offset of $\sim200$ km s$^{-1}$ from the coherent velocity of the disk, meaning that it does not corotate with the starburst disk.
[C~{\sc ii}]-2 is also located along the kinematic minor axis while [C~{\sc ii}]-1 is located along the major axis (Figure \ref{fig;map}).
The spatial and velocity offset suggest that [C~{\sc ii}]-2 is falling into COSMOS-AzTEC-1 from a direction perpendicular to the disk rotation.

We measure the \cii flux of the off-center components in the velocity range where the emission is detected at above $2\sigma$ (Figure \ref{fig;spe}).
In the low-resolution velocity-integrated maps, the peak flux is $0.147\pm0.017$ Jy km s$^{-1}$ beam$^{-1}$ ($8.6\sigma$) for [C~{\sc ii}]-1 and $0.186\pm0.022$ Jy km s$^{-1}$ beam$^{-1}$ ($8.4\sigma$) for [C~{\sc ii}]-2.
On the other hand, the peak flux in the high-resolution maps decreases to $0.087\pm0.021$ Jy km s$^{-1}$ beam$^{-1}$ ($4.2\sigma$) for [C~{\sc ii}]-1 and $0.113\pm0.026$ Jy km s$^{-1}$ beam$^{-1}$ ($4.4\sigma$) for [C~{\sc ii}]-2, indicating that 
these components are spatially extended.
Therefore, we refer to these off-center components as a subcomponent rather than a clump.
We derive 0.17$\pm$0.04 Jy km s$^{-1}$ for [C~{\sc ii}]-1 and 0.62$\pm$0.10 Jy km s$^{-1}$ for [C~{\sc ii}]-2 by using CASA/{\tt imfit}, which fits an elliptical gaussian component on the image.
[C~{\sc ii}]-1 and [C~{\sc ii}]-2 contribute 1 \% and 4 \% to the total \cii flux, respectively.
These two \cii subcomponents are not detected in dust emission while two dust clumps are not detect in \cii emission.
This result indicates that there are variations in the ratio of \cii to far-infrared (FIR) luminosity of clumps and subcomponents.
\cii subcomponents tend to have a larger [C~{\sc ii}]/FIR ratio than the clumps identified by dust continuum emission.
In the \cii subcomponents, molecular gas would not be irradiated by a strong far-ultraviolet field produced by massive stars in contrast to the dust clumps and central region of starburst disks \citep{2019ApJ...876....1T,2019ApJ...876..112R}.

\section{Discussion}

Exploiting the high-quality \cii line cube, we have identified two \cii subcomponents in a bright SMG at $z=4.3$, COSMOS-AzTEC-1.
[C~{\sc ii}]-2 has a large velocity offset of 200 km s$^{-1}$ from the primary disk component while the peak velocity of [C~{\sc ii}]-1 is consistent with that of the disk.
The escape velocity from the galaxy is expected to be $\sim$500 km s$^{-1}$ at the position of [C~{\sc ii}]-2, given that the [C~{\sc i}]-based gas mass of $7.2\times10^{10}~M_\odot$ is mostly enclosed inside a radius of 2.5 kpc \citep{2018Natur.560..613T}.
[C~{\sc ii}]-2 could therefore come back to the galaxy even if it is temporarily going away. 
If assuming that that \cii flux ratios reflect mass ratios between the subcomponents and the starburst disk, 
[C~{\sc ii}]-2 would be at least 10 times less massive than the disk component, which is also supported by non-detection of dust continuum emission.
The physical separation between the primary disk component and [C~{\sc ii}]-2 is 2 kpc, which has not been probed by previous low-resolution (over kpc scales) observations of gas.
\cite{2018ApJ...861...43P} have also identified a \cii subcomponent with a close separation of 2 kpc around a  dusty starburst galaxy at $z=5.67$.
As the integration time of our ALMA observations is only 1 hour, it would be desirable to make similar observations in a large sample of unlensed bright SMGs to investigate whether they commonly have such a subcomponent within the disk.

The large velocity offset and the close separation suggest, not only that COSMOS-AzTEC-1 undergoes a gas-rich minor merger, but also that the accreted gas component does not corotate with the disk, which could possibly behave like counter-rotating streams \citep{2015MNRAS.449.2087D}.
Since gas is not collisionless unlike stars, it can effectively lose the angular momentum by dissipative processes.
When a gaseous component with a different angular momentum is accreted into a rotating disk of the host galaxy, it could stimulate violent disk instability (VDI) where the disk is turbulent and highly perturbed, driving gas inflow into the central region of the galaxy (\citealt{2009Natur.457..451D}, \citealt{2014MNRAS.438.1870D}, \citealt{2015MNRAS.450.2327Z}).
We suggest a scenario that the ex-situ non-corotating subcomponent develops the VDI.
The VDI-driven gas inflow can explain the observed central concentration of the CO(4--3) emission with a half-light radius of $R_{1/2}$=1.2 kpc \citep{2019ApJ...876....1T}.
In the central gaseous disk with a high gas mass surface density, in-situ clumps can be formed through standard Toomre instability \citep{1964ApJ...139.1217T} or non-linear VDI \citep{2016MNRAS.456.2052I} and are gravitationally bound against tidal forces and stellar feedback \citep{2018Natur.560..613T}.

At the moment, it is not clear whether the accretion of only one small non-corotating gas component drastically affects the angular momentum and the spatial structure of gas in much more massive galaxies.
We may need to consider the accretion of multiple gas-rich components with different angular momentum.
For low gas fraction cases, multiple minor mergers with 1:50 can significantly change the morphology and the kinematics of massive galaxies \citep{2007A&A...476.1179B}.
It should be verified by numerical simulations whether multiple minor mergers with high gas fraction drive radial gas inflows into the center of galaxies.

On the other hand, our finding of a non-corotating gas component does not necessarily reject the possibility that COSMOS-AzTEC-1 experiences another major merger in the past, which is an important mechanism to explain extreme starbursts with SFR$\sim1000$ M$_\odot$yr$^{-1}$ \citep[e.g.,][]{2013MNRAS.428.2529H,2019MNRAS.485.5631C} and/or clump formation at high-redshift \citep{2019A&A...632A..98C}.
Recent observations discover a group of $\sim$10 gas-rich galaxies within 200--300 kpc around a bright SMG at $z\sim4$ (\citealt{2018ApJ...856...72O}, \citealt{2018Natur.556..469M}, see also \citealt{2011Natur.470..233C}, \citealt{2013ApJ...776...22H}, \citealt{2017A&A...608A..15B}, \citealt{2018MNRAS.479.3879W}, \citealt{2019ApJ...887...55C} for discoveries of gas-rich companions around dusty star-forming galaxies at high-redshift).
In such high-density environments, gas-rich major mergers are naturally expected to occur and then trigger an extreme starburst in the central galaxy of protoclusters.
We do not detect other 1.1 mm continuum sources with $S_\mathrm{1.1 mm}>100~\mu$Jy per 0.\arcsec3 beam within 80 kpc around COSMOS-AzTEC-1 \citep{2019ApJ...876....1T} while there is an overdensity of optically-selected galaxies with photometric redshift of $4.08<z_\mathrm{phot}<4.60$ \citep{2017A&A...597A...4S}.
The absence of 1.1 mm continuum sources does not positively support that previous major mergers drive gas inflows into into the galaxy center.

Studying stellar kinematics is one of the most effective ways to validate the major merger scenario.
Gas kinematics becomes rotation-dominated in both cases of major and minor mergers (\citealt{2006ApJ...645..986R} for numerical simulations, \citealt{2014ApJS..214....1U} for observations of CO lines)
while only major mergers leave a dispersion-dominated stellar component with $V_\mathrm{rot}/\sigma_0<1$ (\citealt{2006ApJ...645..986R} for simulations, \citealt{2006AJ....132..976R} for observations of Ca triplet lines).
Bright SMGs are faint at $<2~\mu$m but relatively bright at $3-4~\mu$m with an AB magnitude of 22.
Integral field spectroscopy with Near Infrared Spectrograph on James Webb Space Telescope \citep{2016A&A...592A.113D} will enable us to spatially resolve the stellar continuum and the absorption lines at kpc-resolution.
Investigating both gas and stellar kinematics through ALMA-JWST synergetic observations will shed light on the physical mechanism responsible for extreme starbursts in the early Universe.


\
\

We are very grateful to the referee for constructive suggestions to improve the paper.
This paper makes use of the following ALMA data: ADS/JAO.ALMA\#2015.1.01345.S, 2017.1.00127.S. ALMA is a partnership of ESO (representing its member states), NSF (USA) and NINS (Japan), together with NRC (Canada) and NSC and ASIAA (Taiwan) and KASI (Republic of Korea), in cooperation with the Republic of Chile. The Joint ALMA Observatory is operated by ESO, AUI/NRAO and NAOJ.
K.T. acknowledges support by Grant-in-Aid for JSPS Research Fellow JP17J04449. 
Data analysis was in part carried out on the common-use data analysis computer system at the Astronomy Data Center (ADC) of the National Astronomical Observatory of Japan.
This research made use of {\tt FAKEOBS}, a software tool developed by the Nordic ALMA Regional Center. The Nordic ARC node is funded through Swedish Research Council grant No 2017-00648.


\section*{Appendix A}
\section*{A comparison between compact and extended array configuration data}

\begin{figure}[!t]
\begin{center}
\includegraphics[scale=1.0]{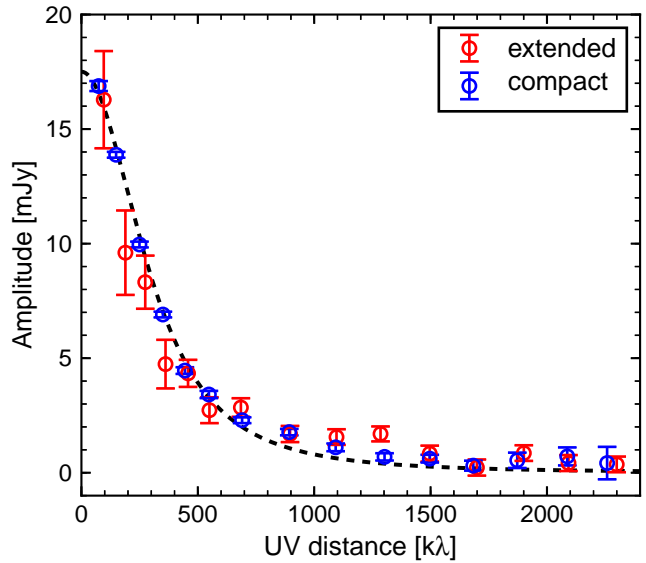}
\end{center}
\caption{
Visibility amplitudes versus $uv$ distances for 860 $\mu$m continuum in COSMOS-AzTEC-1.
Red and blue symbols denote the data in the extended and compact array configurations, respectively.
We overlay an exponential disk model with $S_{860}=$17.5 mJy and a circularized half-light radius ($R_{1/2,\mathrm{circ}}=R_{1/2}\sqrt{q}$) of 1.0 kpc as a reference.
}
\label{fig;uvamp}
\end{figure}

We merge two array configuration data with different $uv$ coverage in this work. 
To cross-check the flux scale between two data sets, we calculate the amplitudes of the vector-averaged visibilities as function of $uv$ distance in each array configuration (Figure \ref{fig;uvamp}).
The errors are calculated from the standard deviations of the real and the imaginary part of visibilities in each bin of $uv$ distance.
Although there is no systematic offset between the visibility amplitudes, we see a small discrepancy between both data sets at 100-400 k$\lambda$, corresponding to the physical scale of the starburst disk in COSMOS-AzTEC-1.
For the extended configuration data, the errors may be larger in this range due to the sparse $uv$ coverage.
The compact configuration data, homogeneously covering the $uv$ plane, has an important role in characterizing the spatial extent of the kpc-scale disk.

\section*{Appendix B}
\section*{Simulations of ALMA observations}

\begin{figure}[!t]
\begin{center}
\includegraphics[scale=1.0]{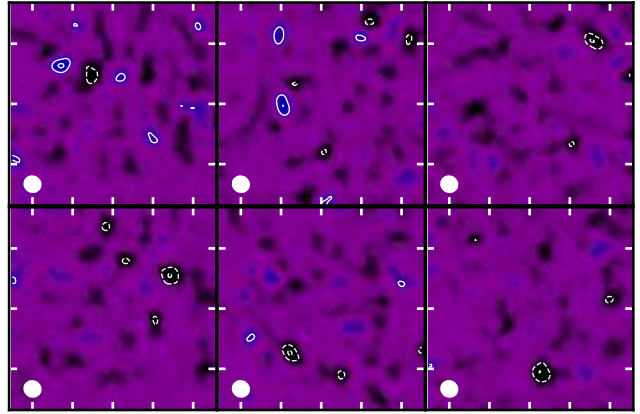}
\end{center}
\caption{
Six simulated residual maps showing positive or negative 5$\sigma$ emission. The image size is 1\arcsec$\times$1\arcsec. The contours are the same as in Figure \ref{fig;dustmap}.
}
\label{fig;sim300_continuum}
\end{figure}

In section \ref{sec;dust_clump}, we identify the residual components after subtracting the best-fit model of a primary disk component in the 860 $\mu$m continuum map.
To evaluate the significance of the residual emission as well as to estimate the realistic uncertainties on the best-fit parameters,
we simulate ALMA observations of 860 $\mu$m continuum observations by using {\tt FAKEOBS}, a software tool developed by the Nordic ALMA Regional Center.
The input model is a smooth disk with S$\acute{\mathrm{e}}$rsic profile, which is derived from the single component fit of the 860 $\mu$m map in section \ref{sec;dust_clump}.
We generate model visibilities with the same uv sampling as the observations, resulting in the same spatial resolution in clean images.
We add noise to visibilities so that the simulated images have the same noise as in the observed ones.
We also made two additional input models with a 20 percent larger and a 20 percent smaller size. 
We run the simulations 100 times with different seed numbers for noise generation in each model.
Then, we fit S$\acute{\mathrm{e}}$rsic models to a total of 300 simulated high-resolution 860 $\mu$m maps by using {\tt GALFIT}.
After subtracting the best-fit model, we search for residual emission within the central region with a radius of 0.5 arcsec. 
Six of 300 simulated residual maps show positive or negative 5$\sigma$ emission (Figure \ref{fig;sim300_continuum}). Therefore, there is 2\% chance that 5$\sigma$ residual emission is artificially created even in a smooth disk. 
The maximum peak by noise fluctuations is 5.3$\sigma$.
We compare between the input and output parameters in the 300 simulations to evaluate the uncertainties.
We obtain the median and standard deviation of $R_{\mathrm{1/2,in}}/R_{\mathrm{1/2,out}}=0.92\pm0.06$, $S_{\mathrm{860,in}}/S_{\mathrm{860,out}}=0.94\pm0.04$, $n_{\mathrm{in}}/n_{\mathrm{out}}=0.91\pm0.06$ and $q_{\mathrm{in}}/q_{\mathrm{out}}=1.02\pm0.03$.
These systematic and random errors are taken into account throughout this work.

We also generate simulated visibilities by using the best-fit models of a disk plus two point sources and a disk plus two Gaussians, shown in Figure \ref{fig;dustmap}. 
From 100 simulations with a disk plus two point sources, we obtain $R_{\mathrm{1/2,in}}/R_{\mathrm{1/2,out}}=0.90\pm0.04$, $S_{\mathrm{860,in}}/S_{\mathrm{860,out}}=0.94\pm0.03$, $n_{\mathrm{in}}/n_{\mathrm{out}}=0.91\pm0.06$ and $q_{\mathrm{in}}/q_{\mathrm{out}}=1.03\pm0.03$ for a disk, $S_{\mathrm{860,in}}/S_{\mathrm{860,out}}=1.04\pm0.13$ for dust clump-2 and $S_{\mathrm{860,in}}/S_{\mathrm{860,out}}=1.05\pm0.11$ for dust clump-3.
From 100 simulations with a disk plus two Gaussians, we obtain $R_{\mathrm{1/2,in}}/R_{\mathrm{1/2,out}}=0.90\pm0.07$, $S_{\mathrm{860,in}}/S_{\mathrm{860,out}}=0.93\pm0.05$, $n_{\mathrm{in}}/n_{\mathrm{out}}=0.90\pm0.08$ and $q_{\mathrm{in}}/q_{\mathrm{out}}=1.03\pm0.05$ for a disk, $R_{\mathrm{1/2,in}}/R_{\mathrm{1/2,out}}=0.98\pm0.11$ and $S_{\mathrm{860,in}}/S_{\mathrm{860,out}}=1.01\pm0.15$ for dust clump-2 and $R_{\mathrm{1/2,in}}/R_{\mathrm{1/2,out}}=1.06\pm0.14$ and $S_{\mathrm{860,in}}/S_{\mathrm{860,out}}=1.09\pm0.14$ for dust clump-3.

\begin{figure*}[t]
\begin{center}
\includegraphics[scale=1.0]{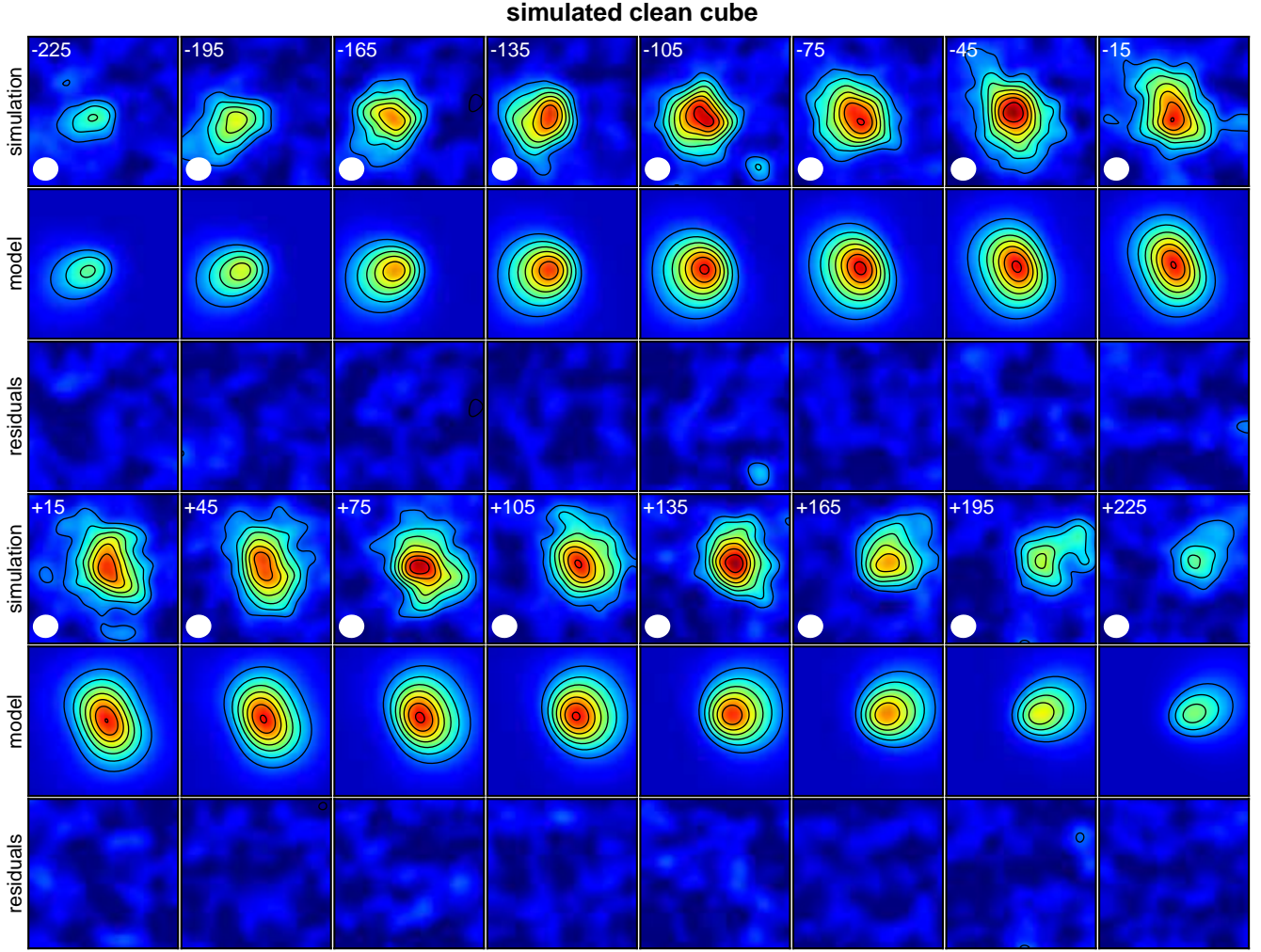}
\end{center}
\caption{
Same as Figure \ref{fig;chmap} but for a simulated clean cube of a smooth rotating disk, which is derived in section \ref{sec;cii_clump}. The contours and the color coding are also same as Figure \ref{fig;chmap}.
}
\label{fig;chmap_sim}
\end{figure*}

In a similar way as the continuum observations, we make CASA simulations for line observations by using smooth rotating disk models with the same parameters derived in section \ref{sec;cii_clump}. 
We make 100 simulated clean cubes with a channel width of 30 km s$^{-1}$.
We also change the size of the input model by $\pm20$ \%, resulting in a total of 300 simulated cubes. 
We fit kinematic models to the simulated cubes by using {\tt Galpak3D} and subtract the best-fit model.
In the 300 simulated cubes within the central region with a radius of 0.5 arcsec, we do not detect residual emission at above $5\sigma$ in the velocity range of $\pm$300 km s$^{-1}$.
The worst case is shown in Figure \ref{fig;chmap_sim}, where the maximum peak is detected at 4.9$\sigma$.
In the wider sky regions with annular radius of 1.0 arcsec and 9 arcsec in the velocity range of $\pm$570 km s$^{-1}$, 5$\sigma$ peaks are artificially created, but the fraction is only 0.004\% of all pixels.
The worst case in the 300 simulated cubes is a negative 6.7$\sigma$ peak.
However, none of them is associated with another 5$\sigma$ peak in the adjacent channel.
This indicates that 5--6$\sigma$ significance in single channel is not a sufficient criterion for blind search of \cii emission although our focus is a small targeted region.
The comparisons between the input and output parameters give $Sdv_{\mathrm{in}}/Sdv_{\mathrm{out}}=0.95\pm0.02$, $V_{\mathrm{max,in}}/V_{\mathrm{max,out}}=0.96\pm0.04$, $\sigma_{\mathrm{0,in}}/\sigma_{\mathrm{0,out}}=0.99\pm0.03$, and $R_{\mathrm{1/2,in}}/R_{\mathrm{1/2,out}}=0.96\pm0.07$.

\section*{Appendix C}
\section*{Analysis of dirty images}

\begin{figure*}[!ht]
\begin{center}
\includegraphics[scale=1.0]{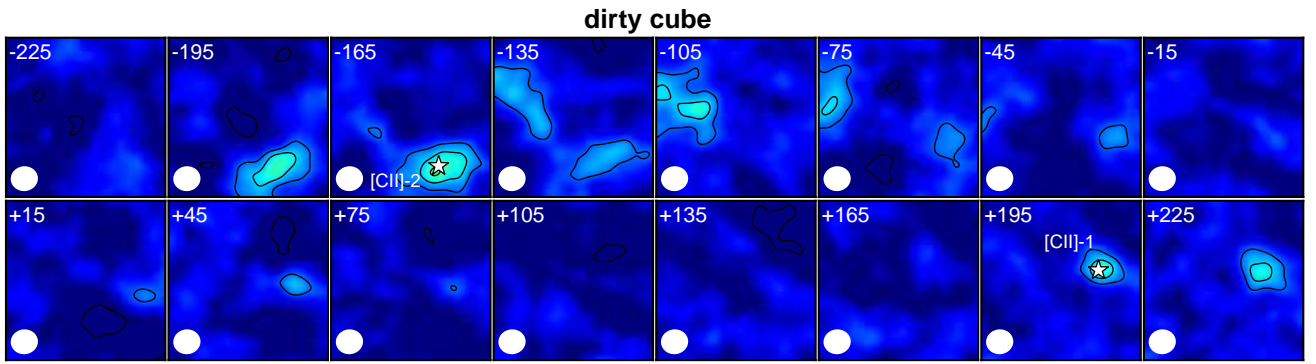}
\end{center}
\caption{
Dirty channel maps after subtracting the best-fit model. The contours and the color coding are also same as Figure \ref{fig;chmap}.
}
\label{fig;chmap_dirty}
\end{figure*}

Another reason to create artifacts of residual emission is deconvolution of the dirty beam in clean process.
Noises could be potentially amplified by deconvolution if the clean is not perfect.
We make the same analysis of a dirty cube to check if the results do not depend on the clean \citep{2019ApJ...882..107R}.
Figure \ref{fig;chmap_dirty} shows residual channel maps of the dirty cube.
Both [C~{\sc ii}]-1 and [C~{\sc ii}]-2 are detected at above $5\sigma$ in two channels, indicating that these residual emission are not enhanced by beam deconvolution in clean.

\end{document}